\documentclass[twocolumn,aps,prl]{revtex4}
\usepackage{amsmath}
\usepackage{graphicx}

\begin{document}

\title{Surface roughness induced electric field enhancement and triboluminescence}
\author{P. Lazi\'c$^{1,2}$ and B.N.J. Persson$^{2,3}$}

\affiliation{$^1$Department of Material Science and Engineering,Massachusetts Institute of Technology, Cambridge, MA 02139, USA }
\affiliation{$^2$IFF, FZ-J\"ulich, 52425 J\"ulich, Germany, EU}
\affiliation{$^3$Division of Nanotechnology and New Functional Material Science,
Graduate School of Science and Engineering,
University of Toyama, Toyama, Japan}

\begin{abstract}
The separation of solids in adhesive contact, or the fracture of solid bodies, often results 
in the emission of high energy photons, e.g., visible light and X-rays. This is believed to be
related to charge separation. We propose that the emission
of high energy photons involves surface roughness and surface diffusion of ions or electrons, resulting in
the concentration of charge at the tips of high asperities, and to electric field enhancement,
which facilitate the discharging process which result in the high energy photons. If the surface
diffusion is too fast, or the separation of the solid surfaces too slow, discharging start
at small interfacial separation resulting in low energy photons. 
\end{abstract}

\maketitle

\pagestyle{empty}

The relative motion between two contacting solids can produce light,
called triboluminescence\cite{review1,review2}. For example, opening an envelope
in a dark room usually result in flashes of blue light from the (pressure sensitive rubber) adhesive interface. 
Recent experiment have shown that
photons with energies up to $\sim 100 \ {\rm keV}$ are produced during the pealing of adhesive tape in 
$10^{-3}$ Torr vacuum. The X-ray pulses were of nano-second duration, produced $\sim 100 \ {\rm mW}$,
and were correlated with stick-slip peeling events.

The origin of triboluminescence is believed to be related to charge separation. In order for charge separation
to generate high energy photons, the discharging process must not occur until the solid walls have been
separated by a relative large distance. If the surface charge density is denoted by $\pm \sigma =\pm n_0 e$ 
(where $e $ is the electron charge and $n_0 $ the ion number density), and if the
charge is uniformly distributed, the voltage drop between the two separating surfaces is given by
$V=E d = 4 \pi \sigma d $ where $d $ is the surface separation. 
The highest energy photons (energy  $\hbar \omega_{\rm max}$) emitted during the
discharging is likely to be $\hbar \omega_{\rm max}= eV= 4 \pi n_0 e^2 d $, and will 
have an energy proportional to the surface separation at the moment of the discharge. In a typical experiment the average
surface charge density $n_0 \approx 10^{14} \ {\rm m}^{-2}$ so that if the discharging would occur at
the separation $d\approx 1 \ {\rm mm}$ one would expect photons with energy up to a few ${\rm keV}$. In the experiment
reported on in Ref. \cite{Camara} photons with energies up to $100 \ {\rm keV}$ were observed, indicating even higher
local charge concentration.  

In vacuum the initiation of the discharging must be related to field assisted emission. That is, such a strong electric
field must be set up at the surface of at least one of the solids that electrons or ions are pulled away from
the surface. This may involve tunneling (mainly for electrons) or thermally induced charge transfer (or 
a combination of both) across or over the barrier towards desorption.
The charged particle is then accelerated by the electric field, and when it hits into the surface of the 
opposite solid it may generate photons (bremsstrahlung) and produce more charged particles, some of which (of opposite sign
as the impacting particle) may accelerate towards the opposite solid and in this way generate a cascade
of charged particles and photons. This may result in very fast (within some nanoseconds) discharging of the surfaces. 

In the normal atmosphere the discharging typically start at lower electric field strength than in vacuum,
resulting in photons with lower maximum energy than in vacuum. The reason for this is most likely the occurrence of
ions in the normal atmosphere which, when entering (or produced by cosmic rays) in 
the space between the charged surfaces, will
be accelerated by the electric field, and trigger the discharging cascade at a lower electric field strength than expected in vacuum.
In fact, one way to avoid charging during separation of solid bodies is to expose the system to high energy
radiation. This creates ions in the surrounding gas 
which are attracted to the surfaces and result in charge neutralization. This method has been
used in some studies of rubber adhesion\cite{Tabor} in order to avoid the electrostatic 
contribution to the work of adhesion, which would result if the surfaces would be charged during 
interfacial separation (see below).

One fundamental problem in triboluminescence is to understand why the discharging typically occurs at relative large
interfacial separation. If the surface charges would be uniformly distributed on the surfaces, then the 
electric field would be constant, independent of the surface separation, at least as long as the surface separation
is large compared to the average distance between two nearby surface charges.
In vacuum the discharging must be initiated by field assisted emission, and if the electric field is not strong enough to
give rise to field assisted emission for short interfacial separation, then one would not 
expect it to occur at large separation either because the magnitude of the electric field strength does not change. 
Of course, quantum mechanical tunneling or thermal activation over a barrier require some average time $\tau $, and if
this time would be long enough (but not too long) 
then this could introduce the delay needed for the surface separation to be so large that
high energy photons would be created.

In the normal atmosphere, if the concentration of ions in the air is low, there could be a 
time-delay before an ion occurs in the thin air film between two separating
surfaces, which could trigger the discharging event.
Thus, in this case the discharging may also occur 
at relatively large interfacial separation, resulting in high energy photons.

It is known that charging during separation of solids decreases with increasing humidity\cite{Harper}. This is probably related
to the formation of thin layers of adsorbed water molecules on the solid surfaces, which results in increased mobility of ions
(and electrons) on the surfaces. Thus, in this case the mobility may be so high that already at small separation
between the solids strong charge concentration occurs, resulting in discharging already at short separation. In addition, if the mobility is high 
enough, ions may move directly from one solid to the other via the area of real contact which exists before complete separation
of the solids. 

\begin{figure}[tbp]
\includegraphics[width=0.47\textwidth,angle=0]{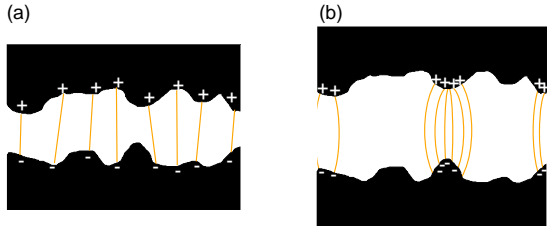}
\caption{
(a) Short time:  During surface separation charges are produced with uniform density on the two surfaces.
(b) Long time: Due to surface diffusion of ions or electrons, the charges will move in such a way as to give rise to an electric
potential $\phi$ which is constant on the surfaces (say $\phi = 0$ and $\phi = V$)
(this state minimizes the total energy). This will result in charge accumulation at the top of high
asperities, giving rise to local electric field enhancement. (Schematic.)}
\label{short.long.time}
\end{figure}

In this paper we suggest another mechanism for the delay in the discharging, which is needed for high energy
photons to be produced. We propose that the discharging involves surface roughness and surface diffusion of the charged particles.
Surface diffusion results in the concentration of charge at the tips of high asperities, and to electric field enhancement
(see Fig. \ref{short.long.time}), which facilitates the discharging process. In this picture, if the surface
diffusion is too fast, or the separation of the solid surfaces too slow, 
discharging would occur already for small interfacial separation resulting in low energy photons.

\begin{figure}[tbp]
\includegraphics[width=0.47\textwidth,angle=0]{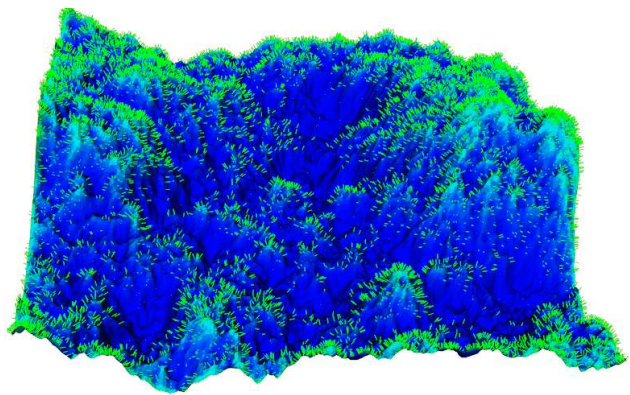}
\caption{
The electric field lines in the vicinity
of a surface with the root-mean-square roughness $=4\ {\rm \mu m}$ and the fractal dimension $D_{\rm f} = 2.2$. Lighter surface colors mean stronger electric field. Upper flat electrode is not shown for clarity.}
\label{onepicfield}
\end{figure}

Almost all natural surfaces and surfaces of engineering interest have surface roughness on many different
length scales, which often can be described as self-affine fractal\cite{fractal,PSSR}.
We have calculated the electric field enhancement for randomly rough (self-affine fractal) surfaces generated mathematically.
The surfaces are assumed to be conducting so that (after long enough time) 
the electric potential is constant on the surface and the electric field
vector is everywhere perpendicular to the surface. 
Surfaces are represented by triangles and the electrostatic problem is solved using the boundary element formulation as implemented 
in the state-of-the-art Robin Hood code \cite{RH}. Only in the boundary element approach one can obtain accurate electric field values, 
especially around spikes. Typical number of triangles used in each calculation was around 200,000. 
Equipotentiality condition at each object  (one rough and one flat surface) in the calculations is satisfied to 4 digits accuracy. 
Electrostatic calculations of such accuracy and with so large number of boundary elements are not feasible 
with any other code today, and in this respect they are the first of its kind.
In the present calculation we have assumed that the two surfaces are oriented parallel to each other. In the experiment  
reported on in Ref. \cite{Camara} the separating surfaces ultimately form an angle of $90^\circ$ but close to the crack edge
we expect more parallel surfaces. If one assume a tilt-angle between the separating surfaces this will generate a non-uniform
(average) surface charge density but if the surface conductivity is not too high, no long-range charge motion can take place and our
simulation results should be at least approximately valid also for the experimental situation of Ref. \cite{Camara}.

In Fig. \ref{onepicfield} we show the electric field lines (short green lines) in the vicinity
of a surface with the root-mean-square roughness $=4\ {\rm \mu m}$ and the fractal dimension $D_{\rm f} = 2.2$.
The surface is a square $L\times L$ with $L= 63 \ {\rm \mu m}$. In order to avoid boundary effects,
in the study presented below we have removed boundary strips of width $0.3 L$, and only use the remaining $0.4L\times 0.4 L$ square in
the middle.

\begin{figure}[tbp]
\includegraphics[width=0.45\textwidth,angle=0]{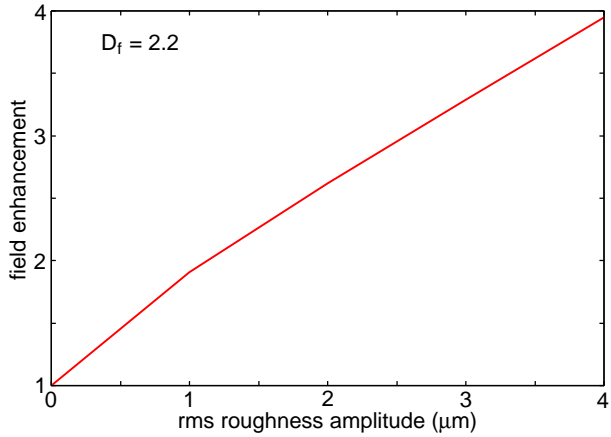}
\caption{
The maximal electric field enhancement for randomly rough surfaces, 
as a function of the root-mean-square surface roughness amplitude.
The fractal dimension $D_{\rm f} = 2.2$ and for a $63 \ {\rm \mu m} \times 63 \ {\rm \mu m}$ square surface unit.
The surfaces have a roll off wavevector $q_{\rm r} =2 q_0= 10^5 \ {\rm m}^{-1}$ and the cut-off wavevector $q_1= 10^7 \ {\rm m}^{-1}$.
The surfaces consist of $200\times 200$ data points. }
\label{rms.enhancement.Df=2.2}
\end{figure}

In Fig. \ref{rms.enhancement.Df=2.2} we show
the electric field enhancement as a function of the root-mean-square surface roughness amplitude.
With field enhancement we refer to the ratio between the strongest
electric field (which typically occurs at a sharp and high asperity tips) on the rough surface, divided by the electric field
for flat surfaces with the same average surface charge. 
The surfaces have the fractal dimension $D_{\rm f} = 2.2$, and are fractal-like over nearly two decades in length scale.
The root-mean-square slope $\xi$ of the surfaces with the root-mean-square surface roughness amplitude $1$, $2$, $3$, $4 \ {\rm \mu m}$ are
$\xi = 0.34$, $0.68$, $1.00$ and $1.35$, respectively. The corresponding total (normalized) 
surface areas are $A/A_0 = 1.054$, $1.20$, $1.39$ and $1.61$,
respectively.

\begin{figure}[tbp]
\includegraphics[width=0.45\textwidth,angle=0]{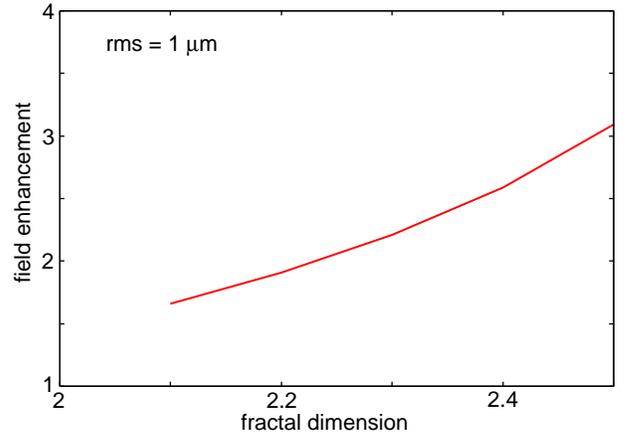}
\caption{
The maximal electric field enhancement for randomly rough surfaces, 
as a function of the fractal dimension $D_{\rm f}$. The root-mean-square surface roughness amplitude $=1 \ {\rm \mu m}$ and
for a $63 \ {\rm \mu m} \times 63 \ {\rm \mu m}$ square surface unit.
The surfaces have a roll off wavevector $q_{\rm r} =2 q_0= 10^5 \ {\rm m}^{-1}$ and the cut-off wavevector $q_1= 10^7 \ {\rm m}^{-1}$.
The surfaces consist of $200\times 200$ data points. }
\label{Df.enhancement.rms=1mum}
\end{figure}

In Fig. \ref{Df.enhancement.rms=1mum} we show the electric field enhancement as a function of the fractal dimension $D_{\rm f}$.
The surfaces have the same size as before, with root-mean-square surface roughness amplitude $=1 \ {\rm \mu m}$.
The root-mean-square slope $\xi$ of the surfaces with the fractal dimension $D_{\rm f}=2.1$, $2.2$, $2.3$, $2.4$ and $2.5$ are
$\xi = 0.27$, $0.34$, $0.43$, $0.57$ and $0.76$, respectively. The corresponding total (normalized) 
surface areas are $A/A_0 = 1.035$, $1.054$, $1.087$, $1.14$ and $1.24$,
respectively. 
Note from Fig. \ref{rms.enhancement.Df=2.2} and \ref{Df.enhancement.rms=1mum} as the roughness amplitude or the fractal dimension increases,
the field enhancement increase monotonically. 

\begin{figure}[tbp]
\includegraphics[width=0.45\textwidth,angle=0]{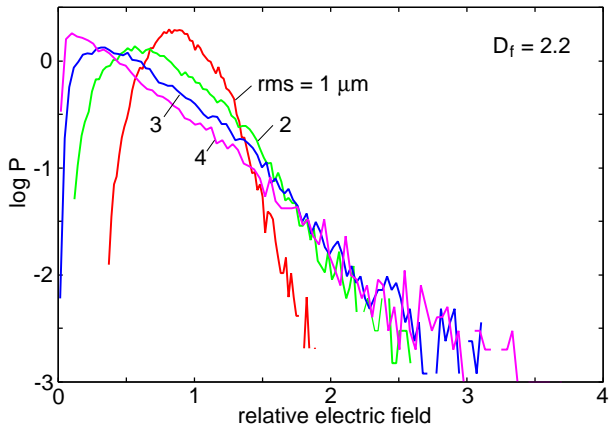}
\caption{
The logarithm (with 10 as basis) of the probability distribution of surface electric field strength for
conducting solids with randomly rough surfaces. 
Results are shown for surfaces with the root-mean-square surface roughness $1$, $2$, $3$ and $4 \ {\rm \mu m}$, and with the
fractal dimension $D_{\rm f} = 2.2$.  For a $63 \ {\rm \mu m} \times 63 \ {\rm \mu m}$ square surface unit.
The surfaces have a roll off wavevector $q_{\rm r} =2 q_0= 10^5 \ {\rm m}^{-1}$ and the cut-off wavevector $q_1= 10^7 \ {\rm m}^{-1}$.
The surfaces consist of $200\times 200$ data points. }
\label{FieldEnhancement.Log10.P.rms=1.a2.a3.a4mum.Df=2.1}
\end{figure}

\begin{figure}[tbp]
\includegraphics[width=0.45\textwidth,angle=0]{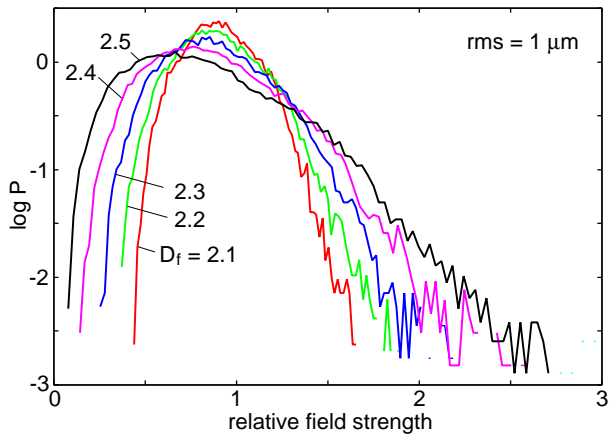}
\caption{
The logarithm (with 10 as basis) of the probability distribution of surface electric field strength for
conducting solids with randomly rough surfaces. 
Results are shown for surfaces with the fractal dimension $2.1$, $2.2$, $2.3$, $2.4$ and $2.5$, and with the
root-mean-square surface roughness $1 \ {\rm \mu m}$.  For a $63 \ {\rm \mu m} \times 63 \ {\rm \mu m}$ square surface unit.
The surfaces have a roll off wavevector $q_{\rm r} =2 q_0= 10^5 \ {\rm m}^{-1}$ and the cut-off wavevector $q_1= 10^7 \ {\rm m}^{-1}$.
The surfaces consist of $200\times 200$ data points. }
\label{FieldEnhancement.Log10.P.rms=1mum.Df=2.1.a2.2.a2.3.a2.4.a2.5}
\end{figure}

In Fig. \ref{FieldEnhancement.Log10.P.rms=1.a2.a3.a4mum.Df=2.1}
we show the logarithm (with 10 as basis) of the probability distribution of surface electric field strength for
conducting solids with randomly rough surfaces. 
Results are shown for surfaces with the root-mean-square surface roughness $1$, $2$, $3$ and $4 \ {\rm \mu m}$, and with the
fractal dimension $D_{\rm f} = 2.2$.  
Similar results are shown in Fig. \ref{FieldEnhancement.Log10.P.rms=1mum.Df=2.1.a2.2.a2.3.a2.4.a2.5}, 
but now as we vary the fractal dimension $D_{\rm f}=2.1$, $2.2$, $2.3$, $2.4$ and $2.5$, while the
root-mean-square surface roughness $ = 1 \ {\rm \mu m}$.  
The electric field strength ($x$-axis) is normalized with the electric field strength for flat surfaces with the same
average surface charge density $n_0 e$. Note that the maximum of the probability distribution shifts to lower field strength as the
roughness amplitude or fractal dimension increases. This result is expected because the average electric field strength
$$\int_0^\infty d E \ E P(E) = 4\pi n_0 e$$
must be constant independent of the roughness, assuming the same average charge density on the surfaces.

It would be interesting to analyze how the statistical properties of the electric field distribution depends on the
number of decades of surface roughness included in the analysis. Such finite-size scaling analysis could allow the
results to be extrapolated to macroscopic systems with roughness on arbitrary number of decades in length scale, which would
be impossible to study directly using numerical methods. Such studies have recently been performed for elastic
contact mechanics between self-affine fractal surfaces\cite{Carlos}, but is beyond the aim of the present study.
Nevertheless, even stronger field enhancements than calculated above may be expected on real surfaces with larger surface area
and with roughness on more decades in length scale than used in out theoretical modeling. 
It is clear that the field enhancement which  arises from surface diffusion of electrons or ions on rough 
surfaces may strongly enhance the probability for field assisted discharging events.

Camara et al.\cite{Camara} have observed X-ray flashes and stick-slip motion in peeling tape in vacuum. 
During peeling in the normal atmosphere the tape motion was instead steady and no X-ray emission was observed,
indicating that in this case the discharging occurred already at much smaller surface separation.
During peeling in vacuum, the peeling force increased approximately linearly with time until discharging occurs, at which point
the pull-force dropped to the value exhibited during peeling in the normal atmosphere. 
The average force necessary for peeling in vacuum was about  $10\% $ higher than during peeling in the normal
atmosphere, and this increase in the peeling force (and a corresponding increase in the work of adhesion) can be attributed
to the extra (electrostatic) work which is done during the charge separation; this is a beautiful illustration
of how important the contribution from electrostatic charging can be to the work of adhesion. We note that in peeling
tape the work of adhesion is very large\cite{large} owing to the highly dissipative processes which occur in the crack tip
process zone (and further away) for soft rubber compounds\cite{Crack}. For stiffer rubber (more cross-linked) the work of adhesion
(in the absence of charging) may be $10^3$ or $10^4$ times smaller, and it is clear that if a similar charging would occur 
in these cases, it would dominate the work of adhesion. 

From the increase in the peeling force due to the charging of the surfaces one can deduce the average surface charge density in the experiments
reported on in Ref. \cite{Camara}. Peeling a distance $L_x \approx 6 \ {\rm mm}$ (the average distance between the 
discharging) gave an increase in the force by about $\Delta F \approx 0.2 \ {\rm N}$. The width of the tape was $L_y= 2 \ {\rm cm}$, and
assuming uniformly distributed charges we get $\Delta F = e E n_0 L_x L_y$. Since the electric field $E=4\pi n_0 e$ we get
$n_0 = [\Delta F /(4\pi e^2 L_x L_y)]^{1/2} \approx 10^{15} \ {\rm m}^{-2}$. If the surface separation at the moment of discharging is
$d \approx 1 \ {\rm mm}$ (which is an upper limit), the highest energy photons would have the energy $\approx 20 \ {\rm keV}$, which is
5 times smaller than observed. Thus, it is clear that charge accumulation must occur on the surfaces not only to
trigger the discharging, but also to generate the high local electric fields necessary to explain the observed maximum photon energy.

To summarize, we have proposed that the emission
of high energy photons involves surface roughness and surface diffusion of electrons or ions, resulting in
the concentration of charge at the tips of high asperities, and to electric field enhancement,
which facilitates the discharging process which result in the high energy photons. If the surface
diffusion is too fast, or the separation of the solid surfaces too slow, discharging start
at small interfacial separation resulting in low energy photons. We have shown that enhancement of the 
electric field is necessary not just for triggering the avalanches but also for
explaining the observed maximum in the emitted photon energy observed in Ref. \cite{Camara}.

\vskip 0.2cm
{\bf Acknowledgments}

B.N.J.P. was supported by Invitation Fellowship Programs for Research in Japan from
Japan Society of Promotion of Science (JSPS).
This work, as part of the European Science Foundation EUROCORES Program FANAS, was supported from funds 
by the DFG and the EC Sixth Framework Program, under contract N ERAS-CT-2003-980409.


\begin{thebibliography}{99}
\bibitem{large}
The work of adhesion $w= F/L_y$, where $F$ is the pull-force and $L_y$ the width of the tape.
In the study in \cite{Camara}, $F\approx 1.6 \ {\rm N}$ and $F_y \approx 2 \ {\rm cm}$ giving
$w\approx 80 \ {\rm J/m^2}$. This is $\sim 3000$ times larger than the work of adhesion 
($w \approx 0.03  \ {\rm J/m^2}$) which would result if all the energy necessary to propagate the 
interfacial crack would result from the energy to
break the weak Van der Waals bonds between the glue molecules and the polyethylene backing.
\end{thebibliography}
\end{document}